\documentclass[showpacs,preprintnumbers,amsmath,amssymb]{revtex4}

\usepackage{graphicx}
\usepackage{dcolumn}
\usepackage{bm}
\usepackage{natbib}
\usepackage{color}

\begin{document}


\title{The Weibull - log Weibull transition of interoccurrence times \\ for synthetic and natural earthquakes}

\author{Tomohiro Hasumi$^1$}
 \email{t-hasumi.1981@toki.waseda.jp}
\author{Chien-chih Chen$^2$}
 \email{chencc@earth.ncu.edu.tw}
\author{Takuma Akimoto$^1$}
 \email{akimoto@aoni.waseda.jp}	
\author{Yoji Aizawa$^1$}
 \email{aizawa@waseda.jp}
\affiliation{$^1$Department of Applied Physics, Advanced School of Science and Engineering, Waseda University, Tokyo, Japan\\ 
$^2$Department of Earth Sciences and Graduate Institute of Geophysics, National Central University, Jhongli, Taiwan}

\date{\today}

\begin{abstract}
We have studied interoccurrence time distributions by analyzing the synthetic and three natural catalogs of the Japan Meteorological Agency (JMA), the Southern California Earthquake Data Center (SCEDC), and Taiwan Central Weather Bureau (TCWB) and revealed the universal feature of the interoccurrence time statistics, Weibull - log Weibull transition. 
This transition reinforces the view that the interoccurrence time statistics possess Weibull statistics and log-Weibull statistics. 
Here in this paper, the crossover magnitude from the superposition regime to the Weibull regime $m_c^2$ is proportional to the plate velocity. 
In addition, we have found the region-independent relation, $m_c^2/m_{max} = 0.54 \pm 0.004$.
\end{abstract}

\maketitle

\section{introduction}
Statistical properties of time intervals between successive earthquakes, henceforward the interoccurrence times and the recurrence times, have been frequently studied in order to predict when the next big earthquake will happen. 
Previous papers have been focused on the determination of the probability distribution and the presentation of the scaling law, as shown in the works of ~\cite{Bak:PRL2002, Corral:PRL2004, Shcherbakov:PRL2005, Abaimov:PEP2008, Enescu:GJI2008, Matthews:BSSA2002}. 
For instance, the Weibull distribution~\cite{Abaimov:PEP2008}, the exponential distribution~\cite{Enescu:GJI2008}, the Brownian passage time  (BPT) distribution~\cite{Matthews:BSSA2002}, the generalized gamma distribution~\cite{Bak:PRL2002, Corral:PRL2004, Shcherbakov:PRL2005}, and the log normal distribution~\cite{Matthews:BSSA2002} are candidates for the distribution function of interoccurrence times and recurrence times. 
Also, in the stationary regime, a unified scaling law was proposed by Corral~\cite{Corral:PRL2004} and then obtained by analyzing the California aftershock data~\cite{Shcherbakov:PRL2005}.\par

Meanwhile, the time interval statistics have been studied by numerical simulations of earthquake models~\cite{Rundle:Book2000}, because real earthquake data are limited. 
For example, both the conceptual spring-block models~\cite{Abaimov:NPG2007, Hasumi:condmat2008_b} and the sophisticated Virtual California model~\cite{Yakovlev:BSSA2006} show the Weibull distribution of the recurrence times. 
One of the authors (TH) also reported that the survivor function of interoccurrence times in the 2D spring-block model can be described by the Zipf-Mandelbrot type power law~\cite{Hasumi:PRE2007}, which has been observed by Abe and Suzuki~\cite{Abe:PA2005}.
 \par

Very recently, a statistical feature of the interoccurrence times, the Weibull - log Weibull transition, was proposed by analyzing the Japan Meteorological Agency (JMA) catalog~\cite{Hasumi:condmat2008_a}. 
We found that the probability distribution of interoccurrence times can be very well fitted by the superposition of the log-Weibull distribution and the Weibull distribution. 
In particular, the distribution of large earthquakes obeys the Weibull distribution with exponent less than unity indicating that the process of large earthquakes is long-term correlated. 
Our earlier results emphasize that the interoccurrence time statistics basically contain both Weibull and log-Weibull statistics, and as the threshold of magnitude $m_c$ is increased, the predominant distribution could change from the log-Weibull distribution to the Weibull distribution. 
Those statistical properties, including the Weibull - log Weibull transition, can be also found in synthetic catalogs produced by the 2D spring-block model~\cite{Hasumi:condmat2008_b}. 
However, the applicability to other tectonic regions of the Weibull - log Weibull transition remains unsolved.  \par

In this study, we further investigate the interoccurrence time statistics by analyzing the Southern California and Taiwan earthquake catalogs. 
Together with previous results from the JMA and synthetic catalogs, we show the universal Weibull - log Weibull transition obtained in all of the catalogs. 
We also demonstrate that a crossover magnitude, $m_c^2$, between the superposition regime and the pure Weibull regime is proportional to the plate velocity, and at the end of this paper we elucidate its implication in the geophysical sense.  \par

\section{Data and methodology}
\begin{table*}
\caption{\label{table1}List of Earthquake Catalogs}
\begin{center}
\begin{tabular}{c|c|c|c|c|c}
Catalog  Name & Covering Region & Term & Number of Earthquakes & $m_{min} $ & $m_c^0$ \\
\hline
\hline
 JMA~\cite{JMA}  & 25$^\circ$\textendash 50$^\circ$N and 125$^\circ$\textendash 150$^\circ$E & 01/01/2001\textendash 31/10/2007 & 130244 & 2.0 & 2.0  \\
\hline
 SCEDC~\cite{SCEDC}   & 32$^\circ$\textendash 37$^\circ$N  and 114$^\circ$\textendash 122$^\circ$W & 01/01/2001\textendash 31/12/2007  & 10838 & 0.0 & 2.0  \\
\hline
 CWB~\cite{CWB}   & 21$^\circ$N\textendash 26$^\circ$N  and 119$^\circ$\textendash 123$^\circ$E & 01/01/2001\textendash 31/12/2007  & 148155 & 0.0 & 2.0  \\
\hline
 Synthetic   & $50\times 50$ (system size) & \textendash & 297040 & 0.0 & 0.3  \\
\hline
\end{tabular}
\end{center}
\end{table*}

For studying the interoccurrence time statistics we analyzed the JMA~\cite{JMA}, the Southern California Earthquake Data Center (SCEDC)~\cite{SCEDC}, the Taiwan Central Weather Bureau (TCWB)~\cite{CWB}, the synthetic catalogs. 
The information on each catalog is listed in Table~\ref{table1}. 
$m_{min}$ and $m_c^0$ correspond to the minimum magnitude and the minimum cutoff magnitude, respectively. We basically consider real earthquakes with magnitude greater than and equal to 2.0 because smaller earthquakes are strongly incomplete.\par

The synthetic catalog is created by the 2D spring-block model with the velocity-weakening friction law~\cite{Carlson:PRA1991}. 
This model is characterized by five parameters: the stiffness $l_x^2$ and $l_y^2$, the decrement of the friction force $\alpha$, the plate velocity $\nu$, and the difference between the maximum friction force and dynamical friction force. 
We set those parameters at $\l_x^2=1, l_y^2=3, \alpha=3.5, \nu=0.01$, and $\sigma=0.01$ for reproducing the realistic Gutenberg-Richter law~\cite{Hasumi:PRE2007}. 
Event magnitude $m$ in the model is defined as $m = m_0+\log_{10} \left( \sum _{i,j}^n  \delta u_{i,j} \right)/1.5$, where $\delta u_{i,j}$ and $n$ are the total slip at $(i, j)$ and the number of slipping blocks, respectively. 
$m_0$ is set at 0.7, so as to shift $m$ to positive. 
The occurrence time of an event is the time when a block slips for the first time during an event. \par

The $n$th interoccurrence time $\tau_{n}$ is defined as $t_{n+1}-t_{n}$, where $t_{n}$ and $t_{n+1}$ are the occurrence times of the $n$th and $n+1$th earthquakes, respectively. 
The methodology is the same as that used in our previous studies~\cite{Hasumi:condmat2008_a, Hasumi:condmat2008_b}. 
Earthquakes with magnitude $m$ above a threshold magnitude $m_c$ in a studied region were considered. 
Then, we calculated the interoccurrence times and regressed the interoccurrence times data in the time domain of $\tau > h$. 
It should be noted that the spatial division was carried out when we examined JMA data. $h$ are set at 0.5 (days) and 0 for real and synthetic earthquakes, respectively. \par
Our previous works~\cite{Hasumi:condmat2008_a, Hasumi:condmat2008_b} revealed that the probability distribution of interoccurrence times $P(\tau)$ can be described by the superposition of the Weibull distribution $P_w$ and the log-Weibull distribution $P_{lw}$, namely,
\begin{equation}
P(\tau)= p\times P_w + (1-p)\times P_{lw}
\end{equation}
where 
\begin{eqnarray}
P_w(\tau) = {\displaystyle \left(\frac{\tau}{\tau_c} \right)^{\gamma -1} \frac{\gamma}{\tau_c} \exp \left[-\left(\frac{\tau}{\tau_c}\right)^{\gamma} \right]}, \; 
P_{lw}(\tau) = {\displaystyle \frac{(\log (\tau /k))^{\delta -1}}{(\log \tau_c' )^{\delta}} \frac{\delta}{\tau} 
\exp \left[-\left(\frac{\log (\tau /k)}{\log \tau_c'}\right)^{\delta}  \right]}, \nonumber
\end{eqnarray}
$\gamma, \tau_c, \delta, \tau_c'$, and $k$ are constants. 
$p$ is the ratio of $P_w$ divided by $P(\tau)$. 
Obviously, $P(\tau)$ is $P_{lw}$ when $p=0$ and $P_w$ when $p=1$. $k$ is fixed at 0.5. 
The log-Weibull distribution is constructed by the log modification of the cumulative distribution of the Weibull distribution. 
Unity $\delta$ reduces the log-Weibull distribution to the power law distribution.

\section{Results}
\begin{table*}
\caption{\label{table2}The interoccurrence time statistics on Southern California and Taiwan. Note that the rms value calculated from the cumulative distribution of $P(\tau)$ is on order of $[10^{-3}]$. $(\cdot)$ stands for the ratio of $P_w$ to $P(\tau)$.}
\begin{center}
\begin{tabular}{c|c|c|c|c|c}
 & & $P_w+P_{lw}$ & $P_w+P_{pow}$ & $P_w+P_{gam}$ & $P_w+P_{ln}$ \\
Region & $m_c$ &  rms  ($P_w/P(\tau)$) & rms  ($P_w/P(\tau)$) & rms  ($P_w/P(\tau)$) & rms  ($P_w/P(\tau)$)\\
\hline
   & 2.0 & 1.7 $(0.45\pm0.006)$ & 7.1 $(0.44\pm0.02)$ & 25 $(0.97\pm0.05)$ & 15 $(0)$  \\
   & 2.5 & 2.3 $(0.58\pm0.007)$ & 7.2 $(0.72\pm0.01)$ & 17 $(1)$ & 6.0 $(0)$  \\
 California  & 3.0 & 5.3 $(0.79\pm0.02)$ & 6.8 $(0.91\pm0.01)$ & 9.1 $(1)$ & 5.4 $(0.57\pm0.03)$  \\
   & 3.5 & 7.6 $(1)$ & 7.6 $(1)$ &  7.6 $(1)$ & 7.6 $(1)$  \\
   & 4.0 & 23 $(1)$ & 23 $(1)$ & 23 $(0.34\pm1.29)$ & 23 $(1)$  \\
\hline
   & 2.5 & 14 $(0.40\pm0.06)$ & 14 $(0.20\pm0.05)$ & 43 $(0.96\pm0.07)$ & 32 $(0)$  \\
   & 3.0 & 3.4 $(0.49\pm0.02)$ & 7.5 $(0.53\pm0.02)$ & 24 $(0.94\pm0.06)$ & 12 $(0)$  \\
 Taiwan  & 3.5 & 5.0 $(0.64\pm0.02)$ & 8.4 $(0.74\pm0.02)$ & 13 $(0)$ & 6.1 $(0)$  \\
   & 4.0 & 3.5 $(0.68\pm0.02)$ & 7.4 $(0.89\pm0.01)$ &  9.3 $(0.83\pm0.03)$ & 3.4 $(0.40\pm0.03)$  \\
   & 4.5 & 5.8 $(1)$ & 5.8 $(1)$ & 5.8 $(1)$ & 5.8 $(1)$  \\
   & 5.0 & 12 $(1)$ & 12 $(1)$ & 12 $(1)$ & 12 $(1)$  \\
\hline
\end{tabular}
\end{center}
\end{table*}

\begin{figure*}[]
\begin{center}
\includegraphics[width=.9\linewidth]{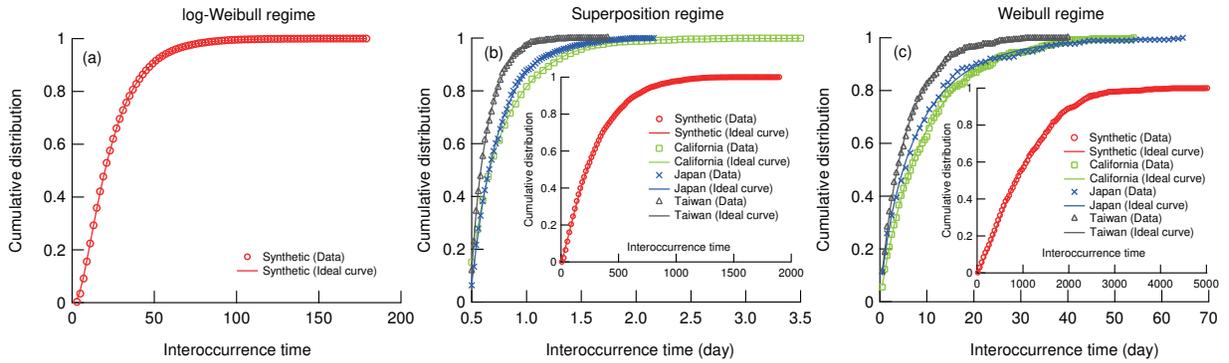}
\end{center}
\caption{The cumulative distribution of interoccurrence times  for different catalogs. (a), (b), and (c) correspond to the log-Weibull regime, the superposition regime, and Weibull regime, respectively. For (b) and (c), the result of the synthetic data are demonstrated in the inset figures.}
\label{data_cumu}
\end{figure*}

The interoccurrence time statistics with different $m_c$ for the SCEDC and TCWB data are studied. 
We use the Weibull distribution $P_w$, log-Weibull distribution $P_{lw}$, power law distribution $P_{pow}$, gamma distribution $P_{gam}$, log normal distribution $P_{ln}$ and their superposition to fit the distribution of the calculated interoccurrence time data. 
As shown in Table.~\ref{table2}, the distribution with larger $m_c$ fairly follows the Weibull distribution. 
Comparing the obtained rms value (please see explanation in the supplementary), we found that $P(\tau)$ is best  fitted by the superposition of the Weibull and log-Weibull distributions. 
The ratio of $P_w$ to $P(\tau)$, $p$, increases as $m_c$ is increased. 
Thus, we can claim that a transition from the Weibull to the log-Weibull transition occurs in both earthquake catalogs of Southern California and Taiwan. 
As an example, together with our previous results for the JMA and synthetic catalogs~\cite{Hasumi:condmat2008_a, Hasumi:condmat2008_b}, we show the cumulative distributions of the interoccurrence times exhibiting the log-Weibull regime, the superposition regime, and the Weibull regime in Figs.~\ref{data_cumu} (a), (b), and (c), respectively. 
Note that Japan data $(\times)$ in Fig.~\ref{data_cumu} is obtained from a divided region spanning 35$^\circ$\textendash 40$^\circ$N and 140$^\circ$\textendash 145$^\circ$E. 
In addition, the pure log-Weibull regime can only be extracted from the synthetic data as shown in Fig.~\ref{data_cumu} (a).

\begin{table*}
\caption{\label{table3}The results of fitting parameters for different distribution functions.}
\begin{center}
\begin{tabular}{c|c|c|c|c|c|c|c|c}
 caption & catalog & $m_c$ & $m_{max} $ & $\gamma$ & $\tau_c$ & $\delta$ & $\tau_c'$ & $p$ \\
\hline
 (a) & Synthetic\footnotemark[1] & 0.7 & 2.8 & $-$ & $-$ & 5.64$\pm$ 0.01 & 51.3$\pm$ 0.06 & 0  \\
\hline
 (b) & Synthetic\footnotemark[1] & 1.6 & 2.8 & 1.15$\pm$ 0.005 & 311$\pm$ 0.73 & 7.36$\pm$ 0.04 & 604$\pm$ 2.04 & 0.69$\pm$0.05 \\
\hline
 (b) & Japan\footnotemark[2]  & 2.5 & 7.2 & 2.91$\pm$ 0.14 & 0.79$\pm$ 0.01 & 1.10$\pm$ 0.03 & 1.43$\pm$ 0.01 & 0.31$\pm$0.02 \\
\hline
 (b) & California & 2.0 & 5.7 & 1.83$\pm$ 0.02 & 0.81$\pm$ 0.006 & 1.13$\pm$ 0.01 & 1.48$\pm$ 0.006 & 0.45$\pm$0.006  \\
\hline
 (b) & Taiwan & 2.5 & 7.1 & 3.35$\pm$ 0.55 & 0.60$\pm$ 0.03 & 1.01$\pm$ 0.09 & 1.18 $\pm$ 0.02 & 0.40$\pm$0.06  \\
\hline
 (c) & Synthetic\footnotemark[1] & 1.6 & 2.8 & 1.29$\pm$ 0.008 & 1115$\pm$ 3.05 & $-$ & $-$ & 1 \\
\hline
 (c) & Japan\footnotemark[2] & 4.5 & 7.2 & 0.82 $\pm$ 0.009 & 7.69$\pm$ 0.07 & $-$ & $-$ & 1 \\
\hline
 (c) & California & 3.5 & 5.7 & 0.95$\pm$ 0.005 & 9.46 $\pm$ 0.04 & $-$ & $-$ & 1  \\
\hline
 (c) & Taiwan & 4.5 & 7.1 & 0.92$\pm$ 0.005 & 5.44$\pm$ 0.02 & $-$ & $-$ & 1  \\
\hline
\end{tabular}
\footnotetext[1]{Dynamical parameters are set at $\l_x^2=1, l_y^2=3, \alpha=3.5, \nu=0.01$, and $\sigma=0.01$.}
\footnotetext[2]{We analyzed the data in the region, ranging from 30$^\circ$N to 35$^\circ$N, and ranging from 140$^\circ$E to 145$^\circ$E.}
\end{center}
\end{table*}

The maximum magnitudes, denoted by $m_{max}$, for the fitting results of $P(\tau)$ in Fig.~\ref{data_cumu} are listed in Table.~\ref{table3}. 
For each studied area $\gamma$ gradually decreases as $m_c$ is increased. 
In the Weibull regime, $\gamma$ obtained from real earthquakes is less than unity, whereas $\gamma$ for our synthetic catalog is greater than unity. 
However, we have noted that $\gamma$ derived from the synthetic data could be also less than 1 by tuning the parameters of the spring-block model. 
For all catalogs, as $m_c$ is increased, $\tau_c$, $\delta$ and $\tau_c'$ increases double exponentially, linearly and exponentially, respectively. 
Particularly, $p$ gradually increases when $m_c$ is increased, indicating that the probability distribution changes as $m_c$ is varied. 
The interoccurrence time statistics exhibit the Weibull - log Weibull transition, which means that the dominant distribution changes from the log-Weibull distribution to the Weibull distribution with increase in $m_c$. 
We thus define crossover magnitudes $m_c^1$ and $m_c^2$ representing magnitudes that the distribution change from the pure log-Weibull regime to the superposition regime and from the superposition regime to the pure Weibull regime, respectively. 
We focus this paper on the values of $m_c^2$ and they are 1.7, 4.0, 3.3, and 4.4 for the synthetic, Japan, California, and Taiwan catalogs, respectively. \par

\begin{figure}[]
\begin{center}
\includegraphics[width=.45\linewidth]{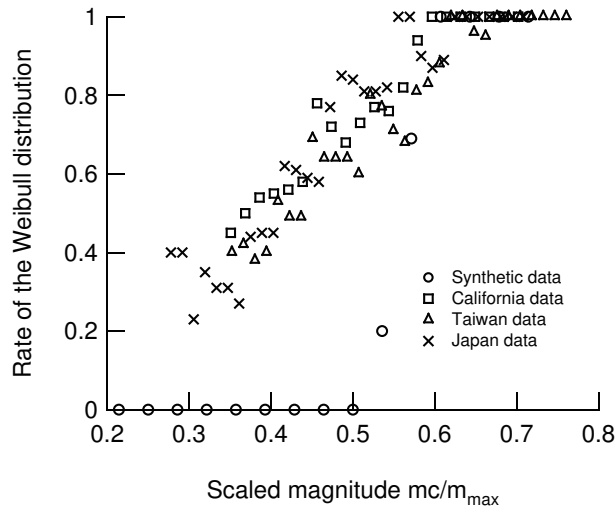}
\end{center}
\caption{The relation between $p$ and the scaled magnitude $m_c^2/m_{max}$. As for the synthetic data, the Weibull - log Weibull transition appears clearly.}
\label{scale_data}
\end{figure}

Most importantly, the universality of the Weibull - log Weibull transition can be clearly shown when we consider a scaled magnitude with $m_c/m_{max}$. 
Fig.~\ref{scale_data} demonstrates the relation between $p$ and $m_c/m_{max}$, the scaled magnitude, for four catalogs. 
As shown in Fig.~\ref{scale_data} $p$ gradually increases with increases in $m_c/m_{max}$, and the scaled crossover magnitude $m_c^2/m_{max}$ of the Weibull - log Weibull transition is estimated to be approximately 0.60 for all of four catalogs. \par

\begin{figure}[]
\begin{center}
\includegraphics[width=.45\linewidth]{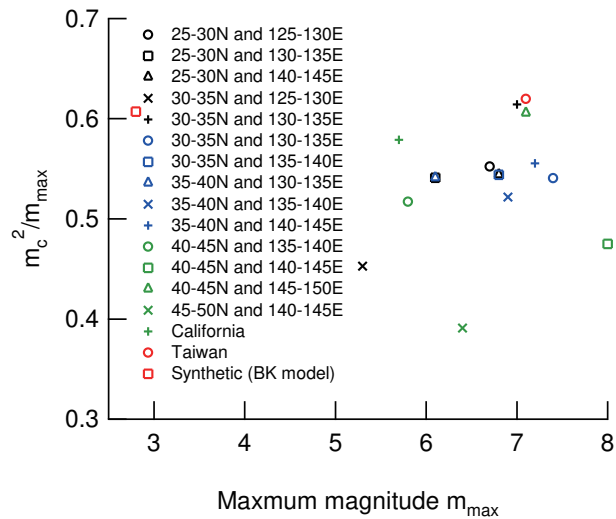}
\end{center}
\caption{The relation between the scaled crossover magnitude $m_c^2/m_{max}$ and the maximum magnitude $m_{max}$ for different regions. $m_c^2$ ranges from 0.39~(45$^\circ$\textendash 50$^\circ$N and 140$^\circ$\textendash 145$^\circ$E) to 0.62 (Taiwan).}
\label{crossover}
\end{figure}

To further discuss the feature of this transition, we summarize our results obtained from 17 different regions (14 regions in Japan and other 3 regions of Southern California, Taiwan, and the synthetic data) in Fig.~\ref{crossover}. 
We surprisingly find an area-independent constant for the scaled crossover magnitude, namely $m_c^2/m_{max} = 0.54 \pm 0.004$. Note that, in Fig. 3, there are three outliers whose values of $m_c^2/m_{max}$ are less than 0.5. 
For those three regions of 30$^\circ$\textendash 35$^\circ$N and 125$^\circ$\textendash 130$^\circ$E, and 45$^\circ$\textendash 50$^\circ$N and 140$^\circ$\textendash 145$^\circ$E, the total numbers of earthquakes are 1739 and 1135, respectively, which are smaller than the earthquake numbers used in other regions (see~Fig. 1 in Ref.~\cite{Hasumi:condmat2008_a}). 
Therefore, the small values of $m_c^2/m_{max}$ are probably caused by the statistical issue of insufficient samplings. 
As for the region of 30$^\circ$\textendash 35$^\circ$N and 125$^\circ$\textendash 130$^\circ$E, $m_{max}$ is 8.0 which is the largest magnitude throughout the JMA catalog we analyzed. 
Therefore, the scaled crossover magnitude for this region tends to be biased small.

\section{Conclusion and discussion}

\begin{table}
\caption{\label{table4}List of the cross-over magnitude and the plate velocity (see in Ref~\cite{Seno:JGR1993, Fowler:1990book}). The notation of PH, EU, PA, and NA represent PHilippine sea plate, EUrasian plate, PAcific plate, and North American plate, respectively.}
\begin{center}
\begin{tabular}{c|c|c|c}
Region & relative plate motion & velocity [mm/yr] & $m_c^2$   \\
\hline
 Taiwan  & PH-EU & 71 & 4.40  \\
\hline
 East Japan  & PA-PH & 49  & 3.80 \footnotemark[1]  \\
\hline
 West Japan  & PH-EU & 47 & 3.78 \footnotemark[2] \\
\hline
 California  & PA-NA & 47 & 3.40  \\
\hline
\end{tabular}
\footnotetext[1]{We take an average using three regions; 25$^\circ$\textendash 30$^\circ$N and 140$^\circ$\textendash 145$^\circ$E $(m_c^2=3.7)$, 30$^\circ$\textendash 35$^\circ$N and 140$^\circ$\textendash 145$^\circ$E $(m_c^2=3.7)$, and  35$^\circ$\textendash 40$^\circ$N and 140$^\circ$\textendash 145$^\circ$E $(m_c^2=4.0)$.}
\footnotetext[2]{We take an average using five regions; 25$^\circ$\textendash 30$^\circ$N and 125$^\circ$\textendash 130$^\circ$E $(m_c^2=3.7)$, 25$^\circ$\textendash 30$^\circ$N and 130$^\circ$\textendash 135$^\circ$E $(m_c^2=3.3)$, 30$^\circ$\textendash 35$^\circ$N and 130$^\circ$\textendash 135$^\circ$E $(m_c^2=4.3)$, 30$^\circ$\textendash 35$^\circ$N and 135$^\circ$\textendash 140$^\circ$E $(m_c^2=4.0)$, and 35$^\circ$\textendash 40$^\circ$N and 135$^\circ$\textendash 140$^\circ$E $(m_c^2=3.6)$.}
\end{center}
\end{table}

We have studied the interoccurrence time statistics of natural and synthetic earthquakes by analyzing the JMA, SCEDC, TCWB, and synthetic catalogs and found the universal Weibull - log Weibull transition of the interoccurrence distribution. 
We emphasize that interoccurrence time statistics contain both Weibull statistics and log-Weibull statistics. 
And, in this paper, we demonstrate the transition does occur for different tectonic settings. 
We also observe the area-independent scaling relation, namely $m_c^2/m_{max} = 0.54 \pm 0.004$. 
Still the origin of the log-Weibull distribution and the Weibull - log Weibull transition remains open. 
Our present work represents the first step to fully understand the interoccurrence time statistics and the Weibull - log-Weibull transition for real earthquakes. \par

Although the scaled crossover magnitudes $m_c^2/m_{max}$ is area-independent, the crossover magnitude $m_c^2$ from the superposition regime to the pure Weibull regime depends on the tectonic region. 
The geophysical implication for the area-dependent $m_c^2$ could be exposed when comparing the plate velocity with {\it averaged} $m_c^2$.  
As clearly seen from Table~\ref{table4}, $m_c^2$ is on the average proportional to the plate velocity. 
That means the magnitude of the largest earthquake, $m_{max}$, for a tectonic region is more or less proportional to the plate velocity since $m_c^2/0.54 = m_{max}$. 
Ruff and Kanamori showed a relation stating that the magnitude of characteristic earthquake which occurs in subduction-zone is proportional directly to the convergence rate~\cite{Ruff:PEPI1980}, which is thus consistent with our results. 
This study therefore suggests a possible physical interpretation for earlier observation about the velocity-dependence of the characteristic earthquake size.

\begin{acknowledgments}
We would like to thank the JMA, SCEDC, and TCWB for allowing us to use the earthquake data. 
The effort of the Taiwan Central Weather Bureau to maintain the CWB Seismic Network is highly appreciated. 
TH is supported by the Japan Society for the Promotion of Science (JSPS) and the Earthquake Research Institute cooperative research program at the University of Tokyo. 
CCC is also grateful for research supports from the National Science Council (ROC) and the Department of Earth Sciences at National Central University (ROC).
\end{acknowledgments}

\end{document}